\documentclass[aps,amsfonts,pre,twocolumn,superscriptaddress,showpacs]{revtex4-1}
\usepackage{epsfig,amsmath,amssymb,bm,epsf,graphics,psfrag,verbatim,subfigure,color}

\def\cstar{c^{*}}
\def\conc{c}

\def\uvec{\mathbf{u}} 

\def \rg{R_g} 
\def \rpart{a} 
\def \force{\mathbf{F}} 
\def \spring{k_\textrm{eff}} 
\def \freq{\omega} 
\def \visc{\eta_f} 
\def \sep{r} 
\def \kb{k_B} 
\def \temp{T} 
\def \modp{G'} 

\begin{document}
\title{Elasticity of colloidal gels: structural heterogeneity, floppy modes, and rigidity}
\author{D. Zeb Rocklin}
\affiliation{Department of Physics, University of Michigan, Ann Arbor, MI 48109}
\affiliation{School of Physics, Georgia Institute of Technology, Atlanta, GA 30332}
\author{Lilian C. Hsiao}
\affiliation{Department of Chemical and Biomolecular Engineering, North Carolina State
University, Raleigh, NC 27695}
\author{Megan Szakasits}
\affiliation{Department of Chemical Engineering, University of Michigan, Ann Arbor, MI 48109}
\author{Michael J. Solomon}
\affiliation{Department of Chemical Engineering, University of Michigan, Ann Arbor, MI 48109}
\author{Xiaoming Mao}
\affiliation{Department of Physics, University of Michigan, Ann Arbor, MI 48109}

\date{\today}

\begin{abstract}
Rheological measurements of model colloidal gels reveal that large variations in the shear moduli as colloidal volume-fraction changes are not reflected by simple structural parameters such as the coordination number, which remains almost a constant. We resolve this apparent contradiction by conducting a normal mode analysis of experimentally measured bond networks of the gels.  We find that structural heterogeneity of the gels, which leads to floppy modes and a nonaffine-affine crossover as frequency increases, evolves as a function of the volume fraction and is key to understand the frequency dependent elasticity.  Without any free parameters, we achieve good qualitative agreement with the measured mechanical response.  Furthermore, we achieve universal collapse of the shear moduli through a phenomenological spring-dashpot model that accounts for the interplay between fluid viscosity, particle dissipation, and contributions from the affine and non-affine network deformation.
\end{abstract}

\maketitle

\emph{Introduction -- } 
Colloidal gels are soft matter with disordered structure and slow dynamics due to short-range, attractive inter-particle forces~\cite{Zaccarelli2007,Trappe2004}.  The attractive interactions stabilize a sample-spanning network of particles.  This network displays mechanical features of a soft solid, including a finite linear elastic modulus at low frequency and the existence of a yield stress at the low shear rate limit~\cite{Bonn2009,Mewis2012}.  
Recent work has established how pair potential interactions and colloidal volume fraction determine the onset of gelation{~\cite{Trappe2000,segre_prl2001,Dinsmore2002,Puertas2004,sciortino2005routes,del2005structure,Dinsmore2006,Lu2008,zaccone_prl2009,colombo2014self}}.  
Observation, by simulation and experiment, of the coincidence of this gel line and the spinodal decomposition boundary suggests a mechanism in which phase instability generates connected regions of high colloidal density.  A prevailing hypothesis is that if the colloidal density of these regions is greater than the glass transition volume fraction, gelation can occur through this heterogeneous mechanism of sequential phase separation and vitrification~\cite{Foffi2005a,Lu2008,royall2008direct}.  Alternatively, attractive interactions of sufficient strength might yield gelation through a homogeneous mechanism in which low-coordination number { (i.e., the number of a particle's contacting neighbors)} networks are stabilized, perhaps only kinetically, through a mechanism such as dynamic percolation~\cite{Dibble2006,Eberle2012}.  Functionally, either gelation mechanism yields a structure in which nearly all particles are spatially localized within a single, sample-spanning network.  

This paper will address the outstanding fundamental question of how such a low volume fraction, disordered, network of the colloid mediates the solid-like rheological properties that are characteristic of gels.  

The low-frequency elasticity of colloidal gels has been predicted from pair potential interactions and microstructure in a few instances.  The linear elastic modulus of fractal cluster gels has been modeled by a microrheological approach, in which the elastic modulus is inversely proportional to the fractal cluster radius and the mean-squared localization length of colloids in the gel.  The localization length can be predicted by summing over the hierarchy of normal modes of the fractal cluster~\cite{Krall1998}.  Mode coupling theory has also been applied to yield the elastic modulus from the localization length in attractive colloidal systems~\cite{Chen2004}, 
albeit with a rescaling required for the effects of voids and clusters in the gel{~\cite{Ramakrishnan2004,kroy2004cluster,zaccone_prl2009,hsiao2014role}}.  A key feature of these theories for the linear elastic modulus is that they connect linear elasticity through two ensemble-averaged quantities, a dynamical localization length and a structural (cluster or particle) scale.  The structural heterogeneity of colloidal gels{, which originates from dynamical arrest and phase separation and plays an important role in the elasticity of gels~\cite{Hsiao2012},} is therefore captured in these models in only a mean-field way.  Moreover, frequency-dependent properties, which would require incorporation of viscous losses, have not been accounted for in these studies.

Here we address these gaps by presenting a theoretical framework to not only compute the frequency-dependent linear viscoelastic modulus of colloidal gels, but also reveal the physics behind the frequency, volume-fraction, and attraction strength dependence of the modulus, as a result of the interplay between floppy modes and mechanical stability, and between affine and nonaffine deformations.  
Our theory includes two parallel approaches to characterize the elasticity of colloidal gels.  The first approach is a microscopic model, in which we take particle positions from measured 3D microstructures of a model colloidal gel and perform normal mode analysis.  Harmonic springs are introduced between neighboring particles with a spring constant extracted from the inter-particle potential in the presence of thermal fluctuations; viscous drag against the affinely-deforming fluid medium is also included in the model.  This microscopic model predicts frequency-dependent shear moduli, showing a crossover from low frequency nonaffine deformations with low rigidity to high frequency affine deformation with high rigidity, in good qualitative agreement with our rheological measurements.  The origin of this crossover is a collection of floppy modes, i.e., particle displacements that don't change bond lengths~\cite{Jacobs1995,lubensky2015phonons,Mao2010,Ellenbroek2011,Zhang2015a,Mao2015,Mao2018}, which are present in colloidal gels as a result of their low coordination numbers and structural heterogeneities.
This observation leads to our second approach, a phenomenological spring-dashpot model based on the Maxwell--Wiechert model of linear solids~\cite{Wiechert1889,Gutierrez-Lemini2016}, incorporating affine and nonaffine limits of deformations and the viscous drag. We obtain good collapse of our experimental shear modulus using this phenomenological model.  Compared to the first approach, this phenomenological model needs no information about the microstructures.  The collapse supports the nonaffine-affine crossover scenario for the frequency dependent shear modulus at different attraction strength and volume fractions.

\emph{Experiment -- } 
The 3D structure of the colloidal gels is studied in conjunction with linear rheological characterization. We synthesize poly(methyl methacrylate) (PMMA) colloids (radius $a=0.58\mu m \pm 4\%$) that are sterically stabilized with a 10-nm layer of poly(12-hydroxystearic acid) (PHSA). The colloids are dyed with fluorescent Nile Red and are suspended at various volume fractions ($\phi=0.15, 0.20, 0.25, 0.30, 0.35,$ and $0.40$) in a mixture of cyclohexyl bromide (CHB) and decalin (63:37 v/v). Non-adsorbing polystyrene (molecular weight MW = 900,000 g/mol, radius of gyration $\rg = 41$ nm) is added at a dilute concentration ($c/c^* = 0.4$ and 0.5, where $c^*$ is the overlap concentration of the depleting polymer) to induce a short-ranged depletion attraction that leads to gelation. Charge screening is provided by adding $1\mu $M tetrabutylammonium chloride (Debye length = 730 nm, zeta potential $\le$ 10 mV) to the gels.

\begin{figure}[h]
\includegraphics[width=.49\textwidth]{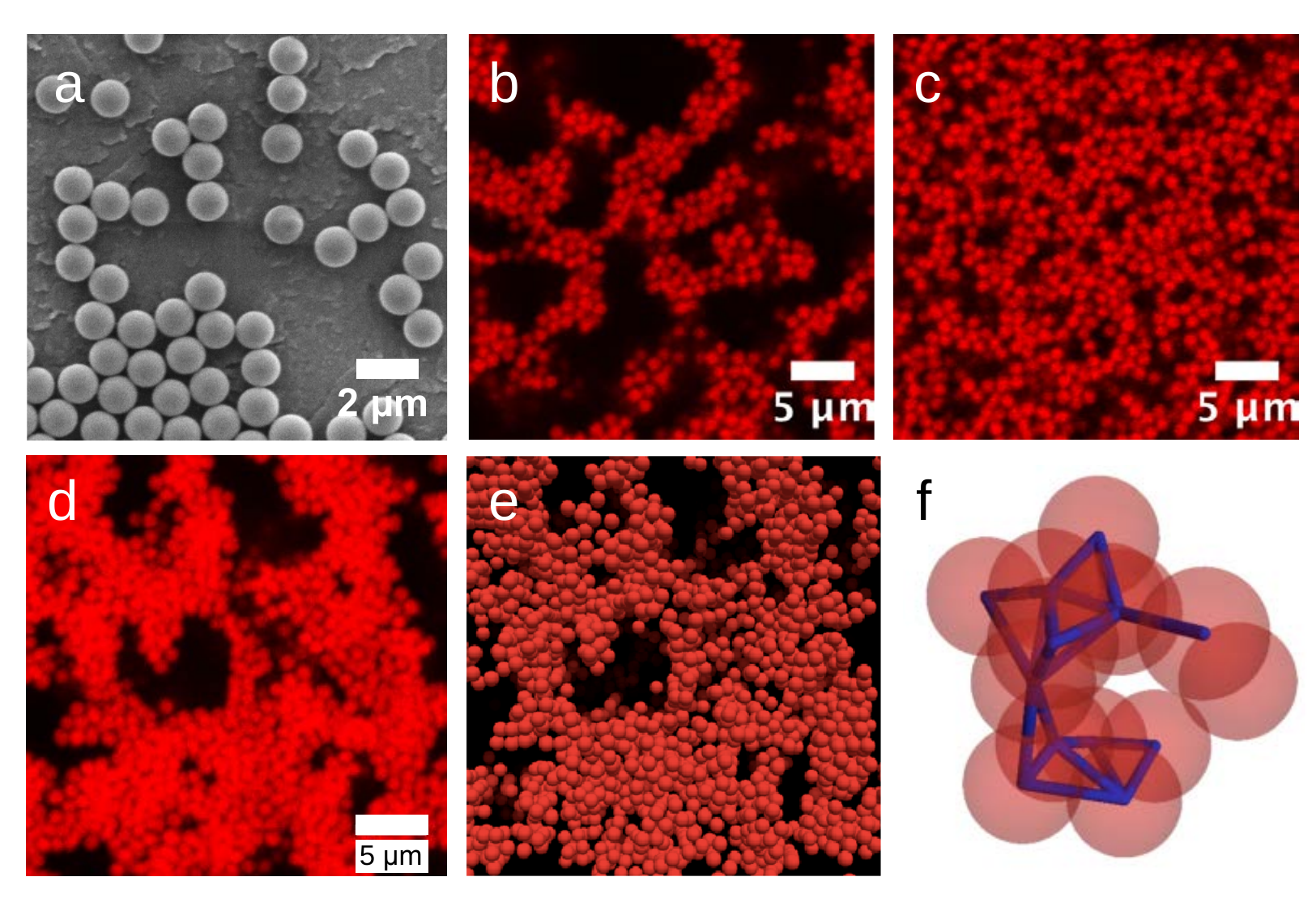}
\caption{(a) Scanning electron micrographs of sterically stabilized poly(methyl methacrylate) colloids 
used to generate gel networks. 
Representative confocal laser scanning microscopy (CLSM) images of colloidal gels at
(b) $\phi = 0.15, c/c^* = 0.4$ and (c) $\phi = 0.40, c/c^* = 0.4$. We collect raw images in 3D and
use image processing to detect particle centroids for theoretical modeling. (d) is a 3D
stack of a gel ($\phi = 0.15, c/c^* = 0.4$) that has been projected onto a 2D plane; (e) shows
a rendering of the microstructure after image analysis. (f) Bonds of attractive contact (blue) between particles
(red) are constructed for particle pairs of distance below $1.5\mu m$.}
\label{FIG:exp}
\end{figure}

\begin{figure}[h]
\includegraphics[width=.5\textwidth]{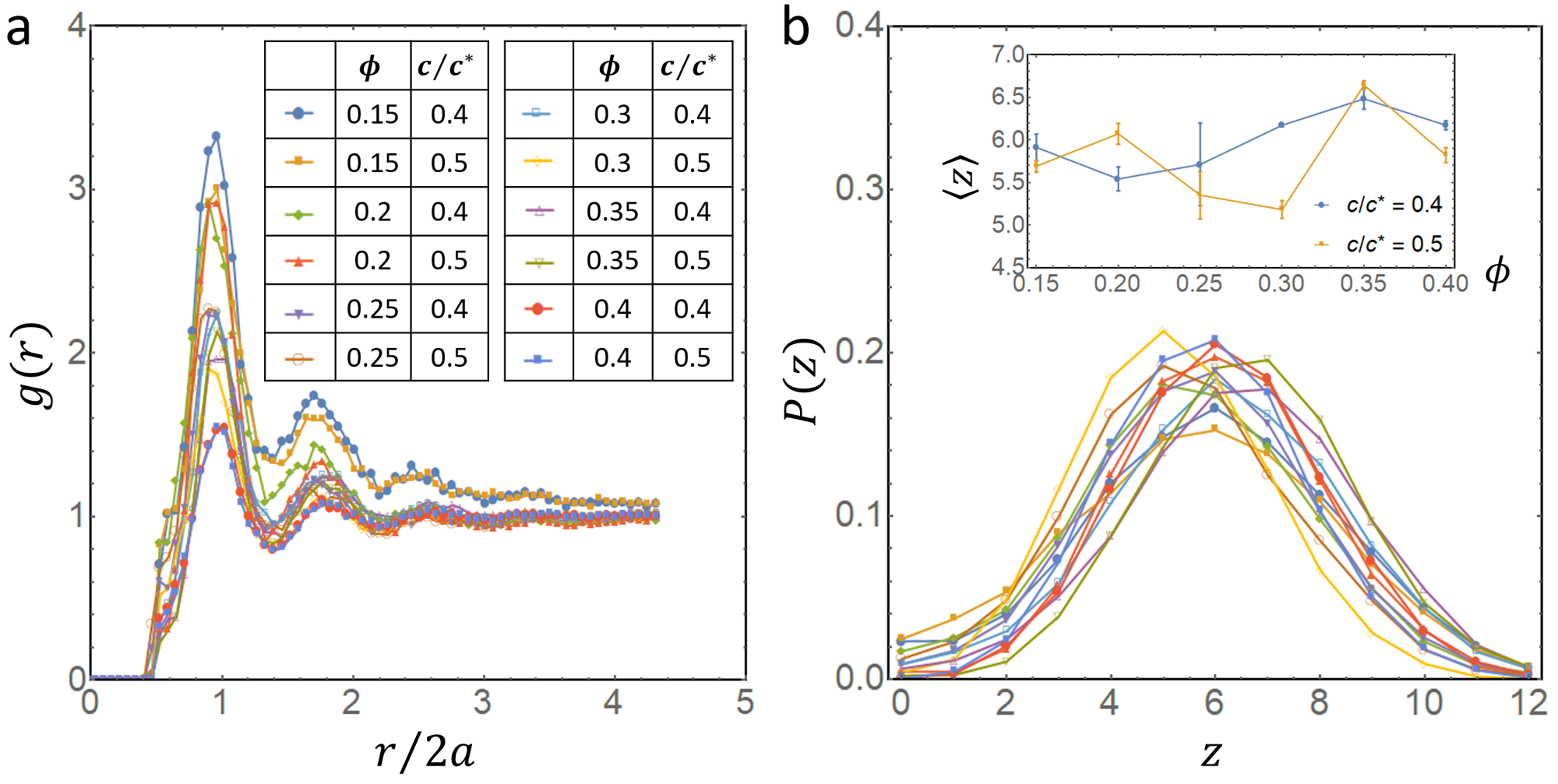}
\caption{(a) Pair correlation function $g(r)$ for each $\phi$ and $c/c^*$. (b) distribution of coordination number, $P(z)$.  Inset shows the mean, $\langle z \rangle$ as a function of $\phi$ at  $c/c^*=0.4,0.5$.
}
\label{FIG:structure}
\end{figure}

Gels are allowed to quiescently equilibrate for 30 minutes prior to imaging and rheological characterization. Figure~\ref{FIG:exp} shows that as $\phi$ increases, the void space of the heterogeneous microstructure is replaced with colloid-rich networks with densely packed, high coordination number regions. 
The confocal microscopy images are obtained from three independent locations within the same sample, at a distance of $\ge 15\mu \textrm{m}$ above the coverslip.  In order to locate particle centroids, we identify using a local regional maximum of intensity in 3D space after smoothing out digital noise in the images~\cite{crocker1996methods}. Fig.~\ref{FIG:exp}(e) is a rendering of the 3D microstructure in Fig.~\ref{FIG:exp}(c), which shows that the 3D structural information used as inputs in our microscopic theory are representative of the gel structure captured from the experiments.

The radial distribution function, $g(r)$, and the coordination number distribution, $P(z)$, are directly computed using the location of the particles in 3D. The $g(r)$ for gels with $c/c^* = 0.4$ and $c/c^* = 0.5$ are plotted in Fig.~\ref{FIG:structure}(a). Particles are considered to be in attractive contact if their separation distance is less that that of the first minimum in the $g(r)$, which is $1.5\mu m$. Figure~\ref{FIG:structure}(b) and inset show that the mean coordination number, $\langle z \rangle$, remains close to six despite the changes in $c/c^*$ and $\phi$. The quantitatively similar nature of the structure for gels with $0.15\le \phi \le 0.40$
is surprising in light of rheological measurements (Fig.~3), in which the low-frequency shear modulus spans more than two orders of magnitude as a function of volume fraction.  As we discuss below, the large variation of shear modulus results from the evolution of the structural heterogeneity, which changes the normal mode structures of the gel, the coupling of which with macroscopic deformations determines the elastic moduli.

\emph{Microscopic Model -- } 
We model colloidal gels as disordered spring networks.  Particle positions are taken from 3D confocal images of the gels, and springs are added to pairs of particles in contact with one another.  Additionally, we treat the fluid medium as moving affinely (i.e., homogeneous deformation field) in response to the external stress, allowing us to approximate the complex effects of the hydrodynamics{~\cite{hoogerbrugge1992simulating,furukawa2010key,vermant2005flow,varga2015hydrodynamics}} as a simple Stokes drag against this affine background{~\cite{Durian1995,Tighe2012,Yucht2013}}. The resulting force on particle $i$ is:
\begin{align}\label{EQ:Force}
\force_i = - \spring \sum_{\langle i,j \rangle}
 \hat{\mathbf{r}}_{ij} \left[\hat{\mathbf{r}}_{ij} \cdot \left(\mathbf{u}_i - \mathbf{u}_j \right)\right] + 6 \pi i \freq \visc \rpart \left( \mathbf{u}_i - \mathbf{u}^{\textrm{aff}}_i\right),
\end{align}
where $ \hat{\mathbf{r}}_{ij}$ is the unit vector pointing from particle $i$ to its bonded neighbor $j$, $\mathbf{u}_i$ is the displacement of particle $i$ from its equilibrium position, $\mathbf{u}^{\textrm{aff}}_i$ is the affine displacement which is $(\Lambda-I) \cdot \mathbf{r}_i$ for a particle with equilibrium position $ \mathbf{r}_i$ under deformation gradient $\Lambda$, 
$\visc =0.0025 \,  \textrm{Pa}\cdot \textrm{s}$ is the fluid viscosity, $\freq$ is the frequency of the driving force, and $\spring$ the effective spring constant.

\begin{figure}[h]
\includegraphics[width=.48\textwidth]{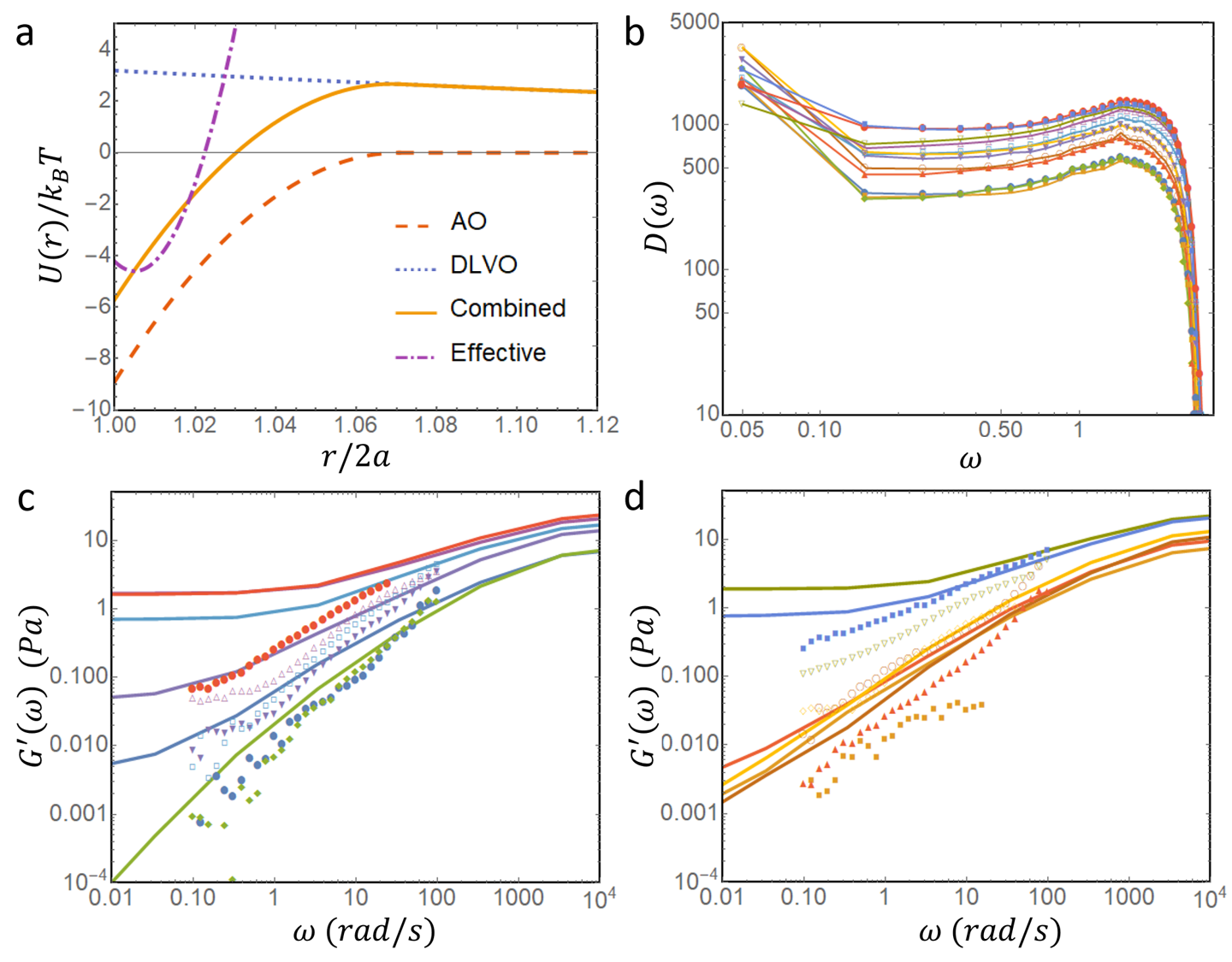}
\caption{(a) Interaction potential between a pair of particles and the harmonic effective potential with spring constant $\spring$ as in Eq.~\eqref{EQ:spring}.
(b) Vibrational density of states calculated from microstructures.  Floppy modes (modes with $\omega=0$) are included in the peak at $\omega=0.05$ (frequency normalization $\sqrt{\spring/m}$ where $m$ is particle mass), which gives the number of modes in the system within $0\le \omega<0.1$.  
(c-d) Experimental (symbols) and model (curves, same color of symbol and curve correspond to the same parameters) values of $G'(\omega)$  at $c/c^*=0.4$ (c) and $c/c^*=0.5$ (d) for different values of $\phi$.  The symbols and colors in (b-d) are the same as in Fig.~\ref{FIG:structure}}.
\label{FIG:modelG}
\end{figure}

In order to obtain the spring constant, we start from the Asakura-Oosawa potential for depletion interaction~\cite{asakura1958interaction},
\begin{align}
U_{\textrm{AO}}(\sep) = - k_B T \frac{c}{c^*}\frac{1}{\rg^3}
\qquad \qquad \qquad \qquad \qquad \qquad
\\ \nonumber
\qquad \qquad
\times \left((\rg+\rpart)^3-\frac{3}{4}(\rg+\rpart)^2 \sep +\frac{1}{16}\sep^3\right),
\end{align}
where $\sep$ is the distance between the centers of the two particles.  Additionally, repulsive electrostatic forces are present, leading to a total potential $U(\sep)$, as described in the SI.
While this interaction is not harmonic, given that the depth of this potential is $\approx 5.7 \, \kb \temp$ ($\conc/\cstar = 0.4$) and $\approx 7.9 \, \kb \temp$ ($\conc/\cstar = 0.5$), thermal fluctuations cause the distance between the pair of particles to explore a significant portion of this potential well.  Thus, instead of taking the curvature of $U(\sep)$ at the minimum, the effective spring constant should be extracted from the thermal fluctuations of the pair distance through a fluctuation-dissipation approach as described in the SI, {resulting in an effective spring constant that accommodates the nonlinearity of the pair interaction over thermal fluctuations in the pair separations:}
\begin{align}\label{EQ:spring}
\spring = \frac{\kb \temp}{\langle \sep^2\rangle - \langle \sep\rangle^2},
\end{align}
where $\langle \sep^2\rangle,  \langle \sep\rangle^2$ are evaluated over all possible bonded separation lengths, $2 \rpart \le \sep \le 2 \left(\rpart + \rg\right)$ for an isolated pair with Boltzmann factor $e^{-U(\sep)/k_B T}$. This results in spring constants of $9.2 \times 10^{-5}$ N/m and $1.8\times 10^{-4}$ N/m for depletant concentrations $c/c^*$ of $0.4$ and $0.5$, respectively. In Fig.~\ref{FIG:modelG}(a) we compare the potentials, with the horizontal and vertical offset of the harmonic effective potential chosen to be the average separation $\langle \sep\rangle$ and the average energy under thermal fluctuations (note that only the curvature of the potential, $\spring$, enters our calculation for shear modulus below).  
A similar calculation was used to estimate the effective spring constant of weakly-aggregated colloidal gels~\cite{Dinsmore2006}.

This model permits the direct calculation of the mode structure of the several thousand particles in the CLSM scan (Fig.~\ref{FIG:modelG}b). We find that in all samples the calculation indicates a significant fraction of collective modes  (on the order $10\%$)  are floppy modes, as a result of the low coordination numbers and the heterogeneous structures~\cite{Jacobs1995,lubensky2015phonons}.  It is worth noting that the existence of these floppy modes does not preclude a finite shear modulus, as the macroscopic shear deformation may not completely overlap with these floppy modes, most of which are localized.  The rest of the vibrational modes exhibit a plateau in the density of states (DOS), sharing similarity to the DOS in jammed packings~\cite{Liu2010}.  

This spring network model recovers the rheological shear response without any free parameters, as shown in Fig.~\ref{FIG:modelG}(c-d). To determine the model's storage modulus $\modp(\freq)$, we subject boundary particles to oscillating shear displacements, allow internal dynamics given by Eq.~\eqref{EQ:Force} and measure the boundary force, as described in more detail in the SI.
We find three regimes of behavior.  
In the high-frequency regime, beyond $\sim 10^4 $ rad/s, Stokes forces dominate, driving the gel to affine displacements and resulting in a plateau in the storage modulus. 
In the low-frequency regime, in contrast, the drag term is negligible and the system is free to assume nonaffine deformations (i.e., spatially varying strain field due to heterogeneity~\cite{Basu2011}),  dominated by the floppy modes, to minimize energy while accommodating the shear boundary conditions.  
For most samples, these boundary conditions cannot be met purely with the floppy modes, resulting in a low-frequency plateau, in accord with previous studies that observed finite elastic moduli even below the isostatic point~\cite{Hsiao2012}.  { The upper limit frequency of this regime is where the fluid drag is comparable to the nonaffine shear modulus, as we discuss more below.  }
In the third regime, which corresponds to intermediate frequencies, shear moduli increase somewhat sublinearly in frequency, displaying nontrivial behavior as nonaffinity is reduced. 
This is the regime accessible via the rheometer, and  good qualitative agreement  is found between direct measurements and the spring model developed from the scans as shown in Fig.~\ref{FIG:modelG}.  
The model (without free parameters) accurately captures the range of moduli observed, the sublinear power-law dependence on frequency and the rough dependence on concentrations of particles and depletant, but falls short of reliable quantitative agreement. 
The discrepancies at low frequencies between microscopic model and experimental data, especially for high density samples,
are due to the fact that the microscopic models are based on microstructures in a small scan window.  The true shear response from rheological measurements at the lowest frequencies involves significantly nonaffine deformations over volumes large compared to the scan window. 
{ Similar types of nonaffine-affine crossovers have been discussed theoretically in the context of disordered spring networks near isostaticity~\cite{Tighe2012,Yucht2013}.}

\begin{figure}[h]
\includegraphics[width=.48\textwidth]{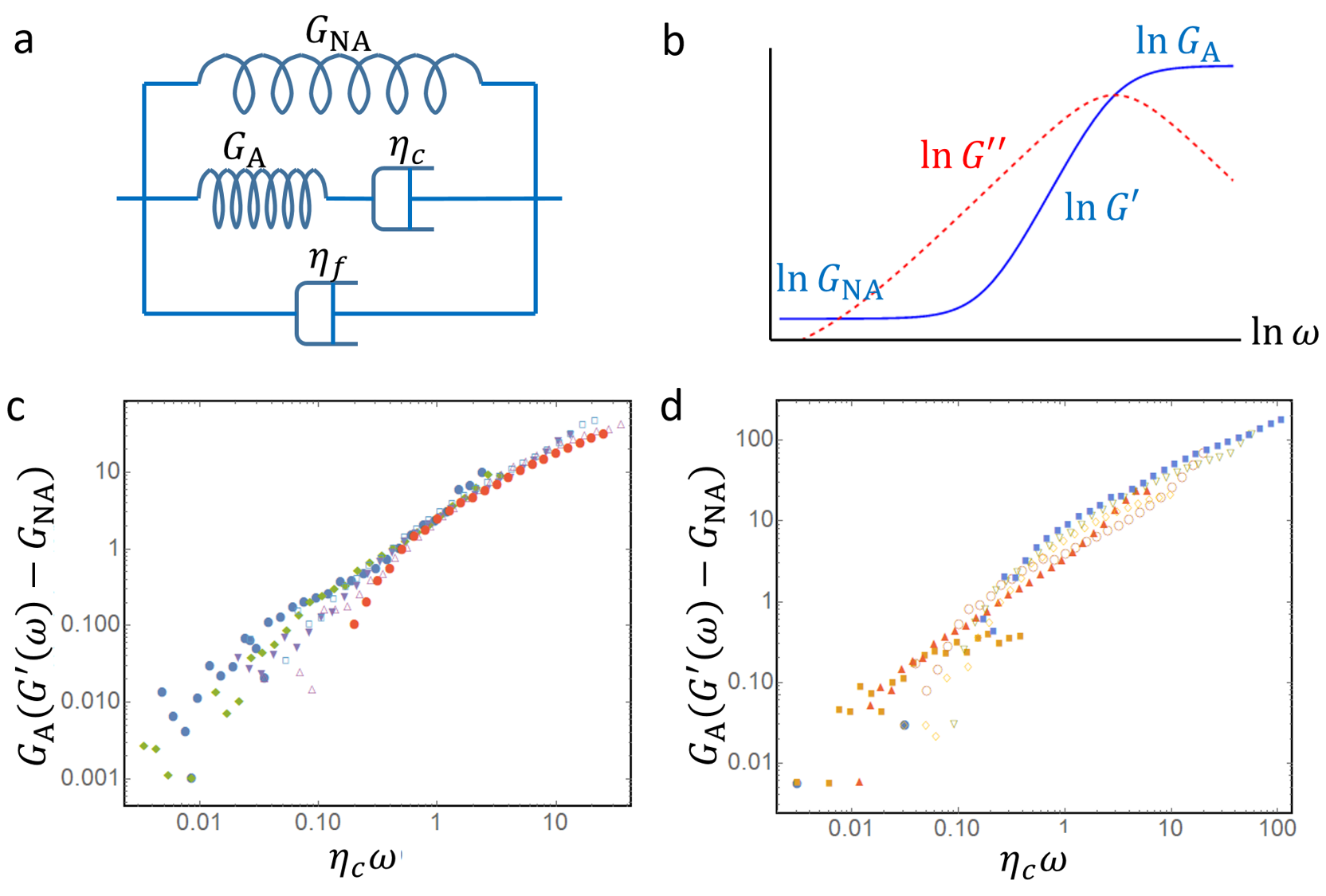}
\caption{(a) Diagram for the phenomenological model. (b) Schematic shear modulus - frequency relation for the model in (a).  (c-d) Collapse of $G'(\omega)$ data using the formula in Eq.~\eqref{EQ:GColl} for $c/c^*=0.4$ (c) and $c/c^*=0.5$ (d).  Colors and symbols are the same as in Fig.~\ref{FIG:structure}. }
\label{FIG:Collapse}
\end{figure}

\emph{Phenomenological Spring-Dashpot Model -- } 
The agreement between our microscopic model and the rheological data suggests a phenomenological picture as shown in Fig.~\ref{FIG:Collapse}(a-b).  The shear response of the gel is a combination of the following effects: at zero frequency shear deformations of the gel are determined by energy minimization which projects the deformation to a collection of the lowest frequency modes, yielding the nonaffine shear modulus $G_{\textrm{NA}}$ (which may vanish at small $\phi$).  We characterize this shear modulus component by a spring of spring constant $G_{\textrm{NA}}$ in our diagram.  
At high frequencies, the fluid which moves affinely drags particles in the gel to move affinely as well, resulting in a much higher shear modulus $G_{\textrm{A}}$.  This increase of shear modulus $G_{\textrm{A}}-G_{\textrm{NA}}$ is a result of fluid drag, so it can be characterized by a spring of spring constant $G_{\textrm{A}}-G_{\textrm{NA}} \simeq G_{\textrm{A}}$ in series with the fluid drag which is characterized by a viscous dashpot of viscosity $\eta_c$ (where the subscript $c$ denote for coupling between fluid and particles).  In addition, in parallel with parts of the diagram described above, there is also the fluid contribution with viscosity $\eta_f$.

Adding up these contributions, the total shear modulus is
\begin{align}\label{EQ:GPheno}
	G (\omega) = G_{\textrm{NA}} 
	+ \left\lbrack G_{\textrm{A}}^{-1}-(i \eta_c \omega)^{-1} \right\rbrack^{-1} -i\eta_f \omega .
\end{align}
Taking the real part of this equation and assuming that $\eta_c \omega \ll G_{\textrm{A}} $,  we have
\begin{align}\label{EQ:GColl}
	G_{\textrm{A}} \lbrack G'(\omega)-G_{\textrm{NA}}\rbrack = (\eta_c \omega)^2
\end{align}
which suggests that our rheological data $G'(\omega)$ can be collapsed into a master straight line.  To obtain this collapse, we use $G_{\textrm{A}}$ from a simple estimate that all bonds are in random orientations in the network (see SI for derivation), 
\begin{align}
	G_{\textrm{A}} \simeq \frac{\phi \langle z \rangle k_{eff}}{10\pi a} ,
\end{align}
and $G_{\textrm{NA}}$ from the low-frequency plateau in $G'(\omega)$ data (computed as average $G'(\omega)$ for $\omega<0.2 \,\textrm{rad/s}$).  We extract $\eta_c$ through the following procedure.  From the imaginary part of Eq.~\eqref{EQ:GPheno} we have
$	G''(\omega) \simeq (\eta_f + \eta_c) \omega $.  With the fluid viscosity $\eta_f$ known, we extract $\eta_c$ for each value of $\phi$ and $c/c^*$ from our $G''(\omega)$ data (fitting $G''(\omega)-\eta_f \omega$ as a straight line for $\omega < 1\, \textrm{rad/s}$ where the linear relation works well).  Using these parameters, we obtain good collapse of $ G'(\omega)$ according to Eq.~\eqref{EQ:GColl}, as shown in Fig.~\ref{FIG:Collapse}c-d.  { It is worth noting that the data collapse into a straight line in the log-log plot as predicted in Eq.~\eqref{EQ:GColl}.  However the slope of the line, instead of $2$, is closer to $1$, indicating perhaps more complicated couplings between the heterogeneous gel structure with the fluid than a simple dashpot.  Related types of scaling collapse of $G(\omega)$, but into a nonlinear master curve, have been discussed in Refs.~\cite{Trappe2000,gardel2004scaling}. In comparison to previous work, we explain the origin of the parameters from the network mechanics and derive them independently rather than fitting from the collapse. }

{ This collapse supports the nonaffine-affine crossover scenario for the frequency dependent shear modulus of colloidal gels, and provides a simple formula to predict gel rheology.}

\emph{Conclusions and Discussion -- }
{ We propose a theoretical framework to understand mechanical properties of colloidal gels, including a method to compute frequency dependent linear shear modulus $G'(\omega)$ from observed microstructures, and a phenomenological spring-dashpot model that collapses $G'(\omega)$ into a master line for different $\phi$ and $c/c^*$.  Our theory is based on analyzing normal modes of the colloidal gel structure as a spring network, which exhibits a large number of floppy modes due to the structural heterogeneity, and  
gives rise to dramatically different static shear moduli at different $\phi$ and $c/c^*$.  The static shear modulus, which involves nonaffine deformations of the network, gives way to a much higher affine shear modulus as a result of viscous drag of the fluid as frequency increases.  The affine shear modulus displays a much smaller spread as a function of $\phi$ and $c/c^*$, because it is not sensitive to the structural heterogeneity.  Our computational model, without any free parameters, accurately predicts the range of moduli observed, roughly how they vary between samples of different particle and depletant densities, and their sub-linear dependence on frequency.

Detailed characterization of the heterogeneous network structure, especially at larger scales which is important in understanding the low frequency shear modulus, and how that affects the gelation transition, as well as how we can control the heterogeneity in experiment and thus tune mechanical response of gels, may be interesting questions in future studies~\cite{Zhang2018}. }

\emph{Acknowledgment -- }
We acknowledge support from NSF under grant number DMR-1609051 (XM), CBET-1232937 (LCH and MJS), and ICAM and Bethe/KIC postdoctoral fellowships (DZR).

\appendix

\section{Interaction potential of the colloidal particles and viscous drag}
The colloidal particles of radius $a = 0.58 \mu\textrm{m}$ are attracted to one another by the depletion effect of the polymers of radius of gyration $R_g = 41 \textrm{nm}$ and concentration $c/c^* = 0.4, 0.5$, leading to the Asakura Oosawa potential~\cite{asakura1958interaction}
\begin{align}
U_{\textrm{AO}}(\sep) = - k_B T \frac{c}{c^*}\frac{1}{\rg^3}
\qquad \qquad \qquad \qquad \qquad \qquad
\\ \nonumber
\qquad \qquad
\times \left((\rg+\rpart)^3-\frac{3}{4}(\rg+\rpart)^2 \sep +\frac{1}{16}\sep^3\right),
\end{align}
Additionally, the colloidal particles undergo a screened electrostatic repulsion of the DLVO form that, neglecting the small van der Waals component, takes the form~\cite{verwey1999theory}
\begin{align}
U_{\textrm{DLVO}}(\sep) = \frac{k_B T Z^2 \lambda_B}{\sep(1+\kappa a)^2} \exp\left[ - \kappa(\sep-2a)\right],
\end{align}
where $\kappa^{-1}$ is the Debye length ($7.3 \times 10^{-7} \textrm{m}$)
and $\lambda_B$ the Bjerrum length ($1.37\times 10^{-8} \textrm{m}$). $Z=108.58$ is the magnitude of the effective charge of the colloidal particles in units of the fundamental charge. Including additionally the van der Waals forces does not appreciably alter the spring constants obtained.

The Stokes drag, treating colloidal particles as ideal spheres with no hydrodynamic interactions, is a force opposing the motion of the particles relative to the fluid, of magnitude
\begin{align}
\mathbf{F}_d = 6 \pi a \eta \mathbf{v},
\end{align}
where $v$ is particle velocity and $\eta = 0.0025 \, \textrm{Pa}\cdot \textrm{s}$ is the dynamic viscosity. Considering harmonic motion of the form $\uvec(t) = \uvec e^{i \freq t}$, $\mathbf{F} = -m \freq^2 \uvec, \mathbf{v} = i \freq \uvec$ (the physical components being the real parts). We neglect the inertial term, which is much smaller than the Stokes term.

\section{Effective spring constant at finite temperature and frequency}
In this section we derive the effective spring constant between bonded pairs of colloidal particles in the gels.  At zero temperature, this would be determined simply by the quadratic term of the expansion about the minimum of the interaction potential between particles shown in Fig.~\ref{FIG:modelG}(a), which would depend largely on the sharpness of the hard-wall repulsion between particles. However, as can be seen in the vertical scale, measured in units of $\kb \temp$, thermal fluctuations drive the particles to explore a broad range of separations over which the potential is quite anharmonic. Despite this nonlinearity in the interaction, the \emph{response}, determined by the free energy, is nevertheless linear for sufficiently small forces and strains (actually, in our case even $U(r)$ contains a statistical contribution, since it includes depletion interactions from polymers). Indeed, this is a particular case of the fluctuation-dissipation theorem, in which the linear response of a thermal system is determined entirely by that system's fluctuations and correlations in the absence of external forces. We now illustrate how this is done in this case.

Consider a particle at temperature $T$ subject both to a one-dimensional potential $U(r)$  and a weak external field $f$ so that  it experiences an effective potential
\begin{align}
U_f(r) = U(r) -  f r,
\end{align}

\noindent and the expectations of observables  is given by

\begin{align}
\langle \hat{O}_f(r) \rangle = \frac{\int d r \, \hat{O}_f(r) \exp \left[ -U_f(r)/k_B T \right]}
{\int d r \, \exp \left[ -U_f(r)/k_B T \right]}.
\end{align}

Following the paradigm of the fluctuation-dissipation theorem, the effective ``spring constant'' is simply defined as the ratio of the applied field to the resultant displacement,
\begin{align}
\spring^{-1} = \frac{\langle r \rangle \rvert_{f} -\langle r \rangle \rvert_{f=0}}{f}.
\end{align}	
We then expand to linear order in the weak field $f$ and obtain the spring constant
\begin{align}
\spring = \frac{k_B T}{\langle r^2\rangle - \langle r \rangle^2} .
\end{align}
Note that this calculation is done using a one dimensional potential.  Extending it to three dimensions will lead to a geometric factor of $O(1)$, but this correction is negligible for small fluctuations which is the case of our interest.

More generally, because the external force of the rheometer is applied at finite frequency, we could consider finite-time correlations (via Fourier transform) to determine a frequency-dependent spring constant. However, since the thermal equilibration time is short compared to the periods explored in the rheometer, the force experienced is effectively constant in time. Thus, to good approximation $\spring(\omega) \approx \spring(0)$, as used in the main text.

This result coincides with that predicted by the equipartition theorem. However, the above approach has the added utility of accounting for nonlinear potentials, demonstrating that the thermal fluctuations root the effective interaction in the long-range features of the electrostatic and depletion interactions rather than the short-range  hard-wall repulsion.

\section{Calculating frequency dependent shear moduli from microstructures}
We obtain positions of particles within the $30.72 \mu \textrm{m} \times 30.72 \mu \textrm{m} \times 25.08 \mu \textrm{m}$ scan window via confocal microscopy.  As discussed in the main text, we model the particle dynamics as the following equation of motion
\begin{align}\label{EQ:Forceapp}
\force_i = - \spring \sum_{\langle i,j \rangle}
 \hat{\mathbf{r}}_{ij} \left[\hat{\mathbf{r}}_{ij} \cdot \left(\mathbf{u}_i - \mathbf{u}_j \right)\right] + 6 \pi i \freq \visc \rpart \left( \mathbf{u}_i - \mathbf{u}^{\textrm{aff}}_i\right).
\end{align}

Within our scan window, we fix the particles within a single bond length of two opposing sides to undergo uniform shear strain at fixed frequencies.  Each type of strain can be written in terms of a deformation gradient $\Lambda$ such that the affine displacement of each particle $\mathbf{r}_i$ is $\mathbf{u}^{\textrm{aff}}_i \equiv (\Lambda-I)\cdot \mathbf{r}_i$.  Boundary particles are assumed to follow the affine displacements.  
We can combine these displacements into a single vector of all the boundary displacements, $\mathbf{u}_B$. 
Internal particles have displacements $\mathbf{u}_I$ which can be nonaffine, such that the total force on each internal particle, as given in Eq.~(\ref{EQ:Forceapp}), is zero (neglecting the small inertial term).
We can use Eq.~(\ref{EQ:Forceapp}) to relate the forces to the displacements via the following matrix equation.

\begin{align}
\label{eq:matrix}
\begin{pmatrix}
    0      \\
    \mathbf{f}_B    
\end{pmatrix}
=
\begin{pmatrix}
    \mathbf{D}_{II}   && \mathbf{D}_{IB}   \\
    \mathbf{D}_{BI}   && \mathbf{D}_{BB}
\end{pmatrix}
\begin{pmatrix}
    \mathbf{u}_I      \\
    \mathbf{u}_B    
\end{pmatrix}
- 6 \pi i \freq \visc \rpart
\begin{pmatrix}
    \mathbf{u}_I -    \mathbf{u}_I^{\textrm{aff}}   \\
    0    
\end{pmatrix},
\end{align}

\noindent where $\mathbf{D}$ is the dynamical matrix from the first term in Eq.~(\ref{EQ:Forceapp}) and we have separated it into boundary and inner parts, and drag force is included as a second term.  
At finite frequencies, this matrix equation is exactly solvable for any given $\mathbf{u}_B$, allowing us to obtain both the nonaffine displacements in the interior $\mathbf{u}_I$ and the forces $\mathbf{f}_B$ on the boundary particles necessary for the shearing motion. At zero frequency, there is a possibility of internal zero modes, leaving the interior displacements not fully defined but not affecting the boundary forces.

Dividing the total boundary force by the area of the boundary, we find the stress induced by the given shear strain. The real and imaginary parts of the ratio of stress to strain are the storage and loss moduli, $G'(\omega), G''(\omega)$. We average over the components of shear moduli obtained from a full Cartesian basis ($\sigma_{xy},\sigma_{yz},\sigma_{zx}$).  
At low frequencies, the nonaffine displacements approach a well-defined limit, which minimizes elastic energy.  This limit is finite for most samples we studied, leading to a low-frequency plateau in $G'(\omega)$, and vanishes only for some low density samples.
As frequency increases, each particle experiences an increasing drag force that inhibits nonaffine deformations, so that the deformation becomes increasingly affine. At high frequencies, the displacements achieve the affine limit, again leading to a (much higher) plateau in the storage modulus.

\section{Affine limit of shear moduli}
In this section we derive the affine limit of shear moduli [Eq.(6) in main text].  Without loss of generality we assume a simple strain with a displacement gradient tensor
\begin{align}\label{EQ:strainlambda}
	\Lambda = 
\begin{pmatrix}
    1   && 0 && \gamma   \\
    0 && 1 && 0 \\
	0 && 0 && 1
\end{pmatrix}
,
\end{align}
and the resulting elastic energy in volume $V$ is 
\begin{align}\label{EQ:Econtinuum}
	E = \frac{1}{2} G V \gamma^2 ,
\end{align}
where $G$ is the shear modulus.

Next we consider this elastic energy as coming from stretching bonds between colloidal particles and obtain a simple estimate for $G$ in the affine limit.  The total number of bonds in volume $V$ for a colloidal gel with volume fraction $\phi$ and mean coordination number $z$ is given by
\begin{align}
	N_{\textrm{bonds}} = \frac{V\phi (z/2)}{(4/3)\pi a^3} ,
\end{align}
where $a$ is the particle radius.  
The elastic energy of one bond (of length $2a$) in direction $(\theta, \psi)$ in polar coordinate, under the affine strain in Eq.~\eqref{EQ:strainlambda} is
\begin{align}
	E_1 = 2 \spring\,  a^2 \gamma^2 \cos^2\theta \sin^2\theta \cos^2 \psi .
\end{align}
Averaging over all solid angles we find
\begin{align}
	\langle E_1\rangle = \frac{2}{15} \spring\, a^2 \gamma^2 .
\end{align}
The total elastic energy in volume $V$ is then the contribution of all bonds, 
\begin{align}
	E = N_{\textrm{bonds}} \langle E_1\rangle .
\end{align}
Equaling the above quantity with the continuum elasticity expression~\eqref{EQ:Econtinuum} we find the affine shear moduli
\begin{align}
	G_{\textrm{A}} \simeq \frac{\phi \langle z \rangle k_{eff}}{10\pi a} ,
\end{align}
which is Eq.(6) in main text.

\bibliographystyle{apsrev4-1}

%

\end{document}